%% file: paper.tex
\documentclass[runningheads]{llncs}
\usepackage{graphicx}
\usepackage{makecell}

\begin{document}
\title{Combining CNNs With Transformer for Multimodal 3D MRI Brain Tumor Segmentation With Self-Supervised Pretraining}

\titlerunning{CNNs With Transformer for 3D MRI Segmentation}

\authorrunning{M. Dobko et al.}

\author{Mariia Dobko\thanks{These authors contributed equally  to the work}
\and Danylo-Ivan Kolinko$^*$ 
\and Ostap Viniavskyi$^*$
\and Yurii Yelisieiev$^*$
}


\institute{The Machine Learning Lab at Ukrainian Catholic University, Lviv, Ukraine 
% \url{https://apps.ucu.edu.ua/en/mllab/} \\
\email{\{dobko\_m,kolinko,viniavskyi,yelisieiev\}@ucu.edu.ua} 
% Avenga, Ukraine \\
% \url{http://https://www.softserveinc.com/en-us} \\
% \email{\{danylo.kolinko,ostap.viniavskyi\}@avenga.com} 
}

\maketitle              % typeset the header of the contribution
\begin{abstract}
We apply an ensemble of modified TransBTS, nnU-Net, and a combination of both for the segmentation task of the BraTS 2021 challenge. In fact, we change the original architecture of the TransBTS model by adding Squeeze-and-Excitation blocks, an increasing number of CNN layers, replacing positional encoding in Transformer block with a learnable Multilayer Perceptron (MLP) embeddings, which makes Transformer adjustable to any input size during inference. With these modifications, we are able to largely improve TransBTS performance. Inspired by a nnU-Net framework we decided to combine it with our modified TransBTS by changing the architecture inside nnU-Net to our custom model. On the Validation set of BraTS 2021, the ensemble of these approaches achieves 0.8496, 0.8698, 0.9256 Dice score and 15.72, 11.057, 3.374 HD95 for enhancing tumor, tumor core, and whole tumor, correspondingly.
Our code is publicly available.\footnote{Implementation is available at \url{https://github.com/ucuapps/BraTS2021\_Challenge}}

\end{abstract}

\input{sections/Introduction}
\input{sections/Methods}

\input{sections/Results}
\input{sections/Conclusion}

\section*{Acknowledgements}

Authors thank Avenga, Eleks, and Ukrainian Catholic University for providing necessary computing resources. We also express gratitude to Marko Kostiv and Dmytro Fishman for their help and support in the last week of competition. 

\bibliographystyle{splncs04}
\bibliography{literature}
\end{document}

%% file: sections/Introduction.tex
\section{Introduction}

% Magnetic Resonance Imaging
% (MRI) is a key diagnostic tool for brain tumor analysis, monitoring and surgery
% planning. Usually, several complimentary 3D MRI modalities are acquired - such
% as T1, T1 with contrast agent (T1c), T2 and Fluid Attenuation Inversion Recover (FLAIR) - to emphasize different tissue properties and areas of tumor
% spread. For example the contrast agent, usually gadolinium, emphasizes hyperactive tumor subregions in T1c MRI modality.

% Glioma is one of the most common type of brain tumors, it begins in the glial cells that surround nerve cells and help them function. Three types of glial cells can produce tumors. It is thus important to classify gliomas by the type of glial cell involved in the tumor and define its genetic features. Correct type diagnosis helps to determine the treatment and predict how the tumor will behave over time.

Glioma is one of the most common types of brain tumors, it can severely affect brain function and be life-threatening depending on its location and rate of growth. Fast, automated, and accurate segmentation of these tumors helps to decrease doctor's time, while also providing a second opinion to make a more confident clinical diagnosis. 
Magnetic Resonance Imaging (MRI) is a well-known scanning procedure for brain tumor analysis. It is usually acquired with several complementary modalities: T1-weighted and T2-weighted scans, Fluid Attenuation Inversion Recovery (FLAIR). T1-weighted images are produced by using short TE (echo time) and TR (repetition time) times, the contrast and brightness of the image are determined by the T1 properties of tissue. In the Flair, sequence abnormalities remain bright but normal CSF fluid is attenuated and made dark, thus Flair is sensitive to pathology. In general, all these modalities indicate different tissue properties and areas of tumor spread.

In this paper, we present our solution for The Brain Tumor Segmentation (BraTS) Challenge, which is held every year and aims to evaluate state-of-the-art methods for the segmentation of brain tumors. In 2021 it is jointly organized by the Radiological Society of North America (RSNA), the American Society of Neuroradiology (ASNR), and the Medical Image Computing and Computer-Assisted Interventions (MICCAI) society. The main task of this challenge is to develop a method for semantic segmentation of the different glioma sub-regions (the "enhancing tumor" (ET), peritumoral edematous/invaded tissue (ED), and the "necrotic" (NCR)) in mpMRI scans. All experiments in this work have been trained and evaluated on BraTS 2021 dataset \cite{data1,data2,data3,data4,data5}.

% Describe annotations

Proposed method is inspired by two approaches: nnU-Net\cite{Isensee2019} and TransBTS\cite{wang2021transbts} model. We incorporate several modifications to their architectures as well as the training process. For example, for TransBTS we add Squeeze-and-Excitation blocks, change the positional encoding in Transformer, incorporate self-supervised pretraining. We also evaluated different postprocessing procedures to improve results and increase generalizability. For instance, we applied connected components, thresholding for noise filtering via class replacements, and Test Time Augmentations (TTA). We use our modified TransBTS to replace a model architecture inside a nnU-Net library, keeping the original preprocessing and training of nnU-Net while also including the useful features of the Transformer in combination with CNN.

%% file: sections/Methods.tex
\section{Methods}
Our solution is based on TransBTS architecture proposed by Wang et al. \cite{wang2021transbts}. However, we also train nnU-Net \cite{Isensee2019} and combine both approaches using ensembling. We also tested the incorporation of our modified TransBTS inside nnU-Net, see Section \ref{architecture_section}.
\par
In the following sections, we describe all of our custom components introduced for data preprocessing, training, and inference postprocessing including architecture modifications.

\subsection{Data Preprocessing and Augmentations}
We have different strategies for training each model that is used for the final ensemble, this includes alterations to data preprocessing.
\par
\textbf{Modified TransBTS:} We combine all MRI modalities of a patient into one 4 channel 3D voxel for the input. The normalization used in our experiments is rescaling according to estimates of the mean and standard deviation of the variable per channel. Every scan is randomly cropped to the shape of 128x128x128. During training we also apply Random Flip for every dimension including depth and Random Intencity Shift according to this formulation:
\[ scale\_factor = Uniform(1.0-factor, 1.0+factor) \]
\[ shift\_factor = Uniform(-factor, factor) \]
\[ image = image * scale\_factor + shift\_factor \]
where factor value was set to 0.1.
\par

\textbf{nnU-Net \& modified TransBTS:} For this training, we use recommended by nnU-Net authors preprocessing which includes per-sample normalization and non-zero mask cropping. The augmentations that were applied during training include Elastic transformation, scaling with a range of 0.85 to 1.25, rotations for all dimensions, gamma correction with a range from 0.7 up to 1.5, mirroring for all axes.

\subsection{Self-pretraining}
Training deep architectures from scratch on 3D data is extremely time-consuming. Transfer learning allows the model to converge faster by incorporating knowledge (weights) acquired for one task to solve a related one. However, the use of any models pretrained on external datasets is forbidden in BraTS Challenge. This is why we perform self-pretraining on the same dataset, with the same model, but for a different task - image reconstruction. We train an autoencoder with an identical encoder from our segmentation model to reconstruct 3D scans. Mean absolute error (MAE) loss was used for this stage.

Since this step is mainly needed to ensure quicker convergence we train the model for 10 epochs. When segmentation starts we load pretrained weights for the encoder part of our TransBTS.

\subsection{Models}\label{architecture_section}
% SEEnBlock & depth change
% TODO: drawing of what was changed compared to transbts
% TODO: Describe TRansBTS architecture in one paragraph, then modifications
TransBTS shows best or comparable results on both BraTS 2019, and BraTS 2020 datasets. The model is based on the encoder-decoder structure, where the encoder captures local information using 3D CNN, these inputs are then passed to Transformer, which learns global features and feeds them to a decoder for upsampling and segmentation prediction. 
\par
The idea behind this architecture is to use 3D CNN to generate compact feature maps capturing spatial and depth information while applying a Transformer following the encoder to handle long-distance dependency in a global space. 

\par
\begin{figure*}[t!]
\begin{center}
\includegraphics[width=10cm,height=5cm]{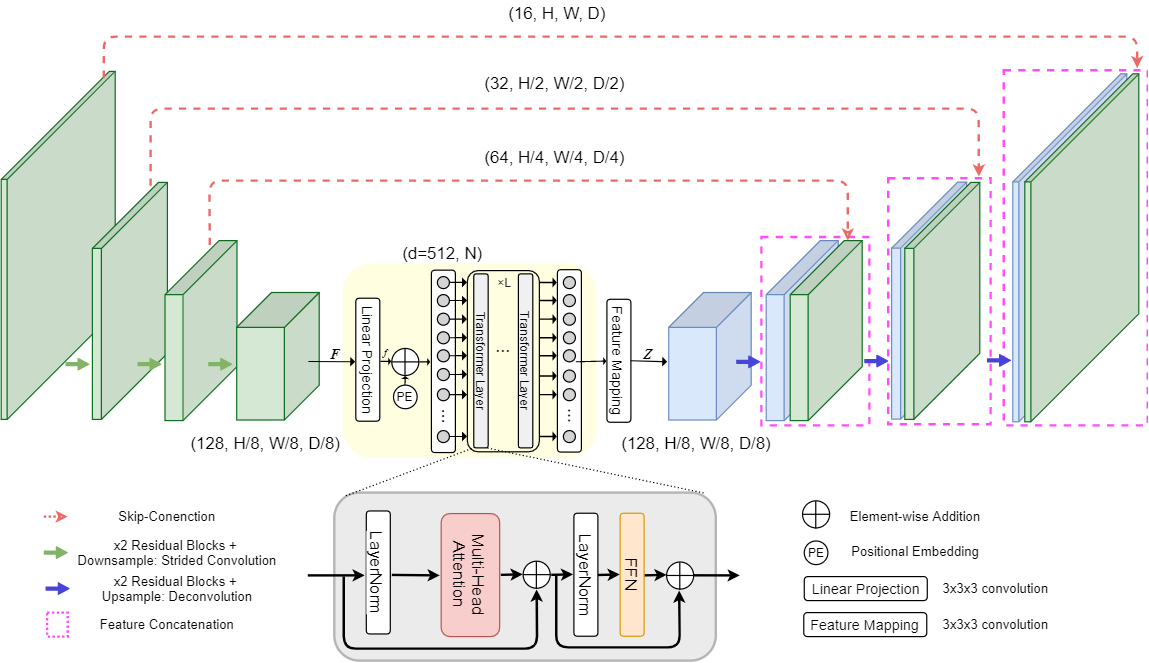}
\end{center}
\vspace{-1.em}
   \caption{Architecture of Original \textbf{TransBTS}, visualization inspired by \cite{wang2021transbts}. Best viewed in color and zoomed in.}
\label{fig:transbts}
\end{figure*}
\begin{figure*}[t!]
\begin{center}
\includegraphics[width=10cm,height=5cm]{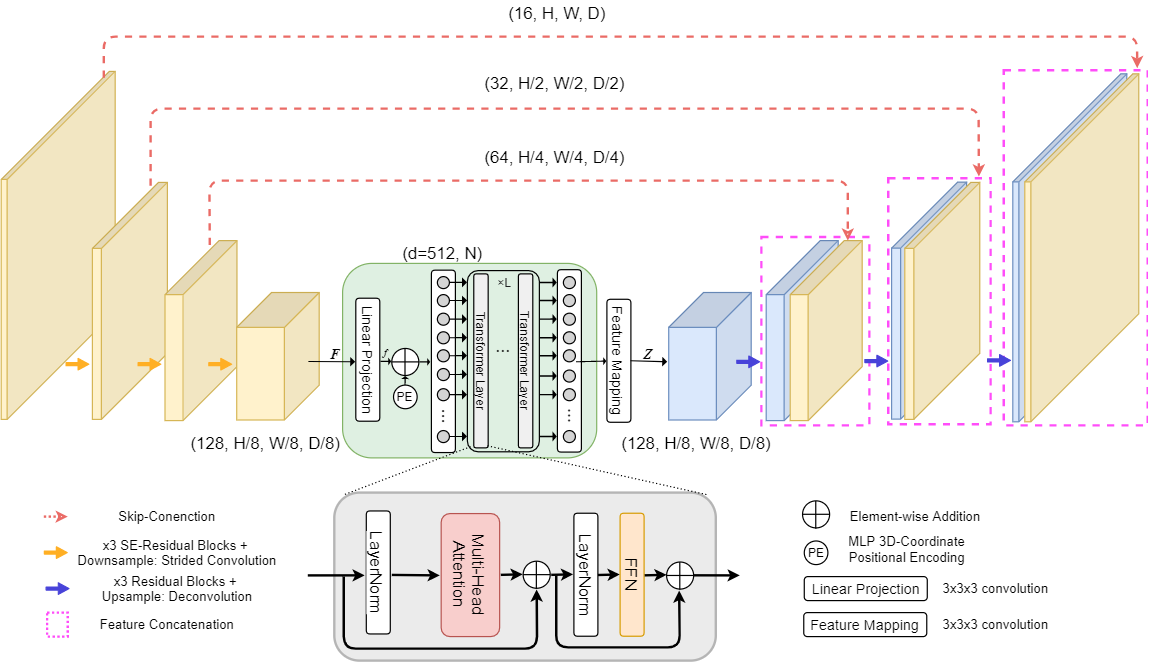}
\end{center}
\vspace{-1.em}
   \caption{Architecture of Our \textbf{Modified TransBTS}, visualization inspired by \cite{wang2021transbts}. Best viewed in color and zoomed in.}
\label{fig:transbtsours}
\end{figure*}

Our custom modifications (see Figure\ref{fig:transbts} and Figure\ref{fig:transbtsours} for comparison with original TransBTS):
\begin{itemize}
  \item We add \textbf{Squeeze-and-Excitation} blocks \cite{Hu2018} to every layer of an encoder. SE blocks help perform dynamic channel-wise feature recalibration.  
  \item The \textbf{depth} of the model was increased compared to TransBTS by adding one layer in encoder and correspondingly in decoder.
  \item We also replaced \textbf{positional encoding} from TransBTS with a learnable MLP block, for more details please refer to Section \ref{mlp}.
\end{itemize} 

nnU-Net\cite{Isensee2019} proposes a robust and self-adapting framework based on three variations of UNet architecture, namely U-Net, 3D U-Net, and Cascade U-Net. The proposed method dynamically adapts to the dataset's image geometry and, more critically, emphasizes the stages that many researchers underestimate. At these stages, a model can get a significant increase in performance. These are the following steps: preprocessing (e.g., normalization), training (e.g., loss, optimizer setting, and data augmentation), inference (e.g., patch-based strategy and ensembling across test time augmentations), and a potential post-processing. It shows that non-model changes in the solution are just as important as model architecture. So we decided to exploit the nnU-Net pipeline to train modified TransBTS.

\subsection{MLP for Positional Encoding}\label{mlp}
\begin{figure}[t!]
\centering
{
\includegraphics[width=10cm,height=2cm]{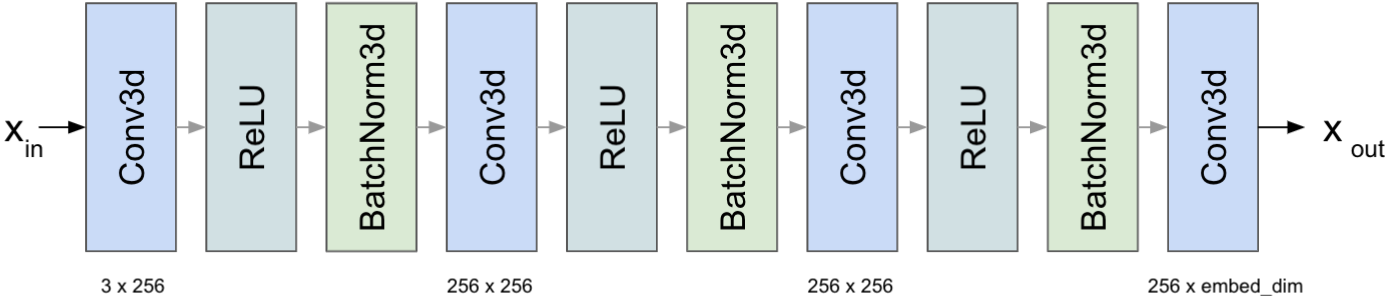}
}
\vspace{-1em}
\caption{
The architecture of MLP 3D-coordinate module for positional embeddings. 
\vspace{-1em}}
\label{fig:mlparch}
\end{figure}

In TransBTS the learnable position embeddings, which introduce the location information, have fixed size. This results in limited input shape for inference since test images can not deviate in scale from the trained set if positional code is fixed. To address this issue we include a data-driven positional encoding (PE) module in form of Multilayer Perceptron (MLP).  
By directly using a 3D Convolution 1x1 we eliminate the problem with fixed resolution and add extra learnable parameters useful for positional embeddings. The MLP architecture consists of three consecutive blocks, where a single block has 3d convolution, relu activation followed by batch norm. This operation is formulated as follows:
% \[ MLP_{b} = BatchNorm3d(ReLU(Conv3d(input))) \]
\[ output = MLP(input) + input \]
where MLP block is displayed in Figure \ref{fig:mlparch}.

\subsection{Loss}
Our training objective for the TransBTS model is the linear combination of Dice and cross-entropy (CE) losses \cite{Isensee2020,Isensee2021} while the original TransBTS model was trained with solely softmax Dice loss. The loss operates on the three-class labels GD-enhancing tumor (ET — label 4), the peritumoral edematous/invaded tissue (ED — label 2), and the necrotic tumor core (NCR — label 1). The best weight between the two loss components was experimentally chosen to be 0.4 for Dice and 0.6 for cross-entropy. 
\par 
When we combine nnU-Net with TransBTS architecture, we also use Dice with CE, but in this case, the weight for both of them is the same, so they have an equal contribution. 
% TODO: Learning rate describe the modifications due to autoencoder pretraining.

\subsection{Optimization}
We apply mixed-precision training, which enables both computational speedup and memory optimization. It is achieved by performing operations in half-precision format and requires two steps: setting the model to use the float16 data type where possible and adding loss scaling to keep small gradient values. 

\subsection{Postprocessing and Ensembling}

First, to reduce memory consumption, we divide the original volume with shape 240x240x155 into eight overlapping regions with the shape 128x128x128. After each part has gone through the model, we form our final prediction by combining all regions. In places where regions intersect, there will be a prediction of the latter. 

Secondly, we perform connected-component analysis (CCA) of ground truth labels on the whole training set. The connected-component analysis applies graph theory, where the input data is labeled based on the given heuristics. The algorithm disassembles the segmentation mask into components according to the given connectivity. CAA can have 4-connected-neighborhood or 8-connected-neighborhood. Eventually, we remove components that are smaller than a 15-voxel threshold. This postprocessing, however, didn't give any severe improvements, so we don't include it in final submissions.

We adopted the idea of another paper \cite{wang2020modalitypairing}, which states that after analyzing the training set, authors noticed that some cases do not have an enhancing tumor class. Therefore, if the number of voxels of this class does not exceed the experimentally selected threshold of 300 pixels in our prediction, we replace it with necrosis.

To increase the performance and robustness of our solution, we create an ensemble of our trained models. One such submission included modified TransBTS trained for 700 epochs and nnU-Net with default configuration trained for 1,000 epochs, in Table \ref{comparison_leaderboard} this experiment is named 'nnU-Net + Our TransBTS'. The weights for probabilities were selected separately for each class (the first coefficient corresponds to nnU-Net, while second to our custom TransBTS): 0.5 and 0.5 for NCR, 0.7 and 0.3 for ED, 0.6 and 0.4 for ET. \par
Our final solution is also an ensemble, which averages probabilities of three models: nnU-Net with default training, nnU-Net trained with our custom TransBTS, and our custom TransBTS trained independently for 700 epochs. The coefficients for these models (following the same order as they were named) are:  0.359, 0.347, 0.294 for NCR class, 0.253, 0.387, 0.36 for ED, 0.295, 0.353, 0.351 for ET class. The results can be viewed in Table \ref{comparison_leaderboard} under the name 'Our Final Solution'.

%  overlapped cropping for inference, 

% Connected component, voxel threshold for removal, 300 pixels in a voxel for 3rd class, we change to 1st

%% file: sections/Results.tex
\section{Results}
We evaluate our proposed method on BraTS 2021 dataset and provide a short ablation study for our postprocessing customization. We compare several combinations of ensembling with training other approaches on the same data.

\subsection{Metrics}
This year's assessment follows the same configuration as previous BraTS challenges. Dice Similarity Coefficient and Hausdorff distance (95\%) are used, a result of aggregation of all of these metrics per class determines the winners. The challenge leaderboard also shows Sensitivity and Specificity per each tumor class.

\subsection{Training phase and Evaluation}
We split training data into two sets: training (1,000 patients) and local validation (251 patients). This is done to have an opportunity to evaluate our customizations locally and see their impact before submitting to the challenge. Many of our additional modifications have shown negative or no impact locally, so they weren't used in the final method. These include but are not limited to 2D segmentation model TransUNet\cite{chen2021transunet}, an ensemble of our 3D model and 2D TransUNet, gamma correction tta, etc.

\par
To see which architecture or ensemble shows the best performance we computed metrics locally and/or on BraTS2021 validation. On local validation TransBTS trained for 500 epochs, for instance, shows 0.56284, 0.85408, 0.8382 Dice for NCR, ED, and ET correspondingly, and 21.858, 4.3625, 3.6681 HD score.
While the model with our modifications trained for a same number of epochs helps us achieve 0.7737, 0.8444, 0.8424 Dice and 5.017, 4.57, 3.125 HD.
Comparison of our solution with other models is displayed in Table \ref{comparison_leaderboard}. We trained nnU-Net with default configurations to analyze our results and proposed model, we also tested the ensemble of our method together with nnU-Net.

\par
For postprocessing, we applied different hyperparameters and evaluated them on local validation. Best configurations were also estimated on leaderboard validation data, refer to Table \ref{postproc_experiments}. We tested several tta combinations on the local validation set to determine the best fit, see results in Table \ref{tta_experiments}. 

\begin{table}[t!]
\centering
\caption{Comparison of different methods on BraTS2021 validation set.  Dice is computed per class, HD corresponds to Hausdorff Distance. nnU-Net + our TransBTS stands for an ensemble (averaging class probabilities) of both models trained separately (our for 700 epochs), 'Our TransBTS inside nnU-Net' is a proposed model based on TransBTS wrapped in nnU-Net training pipeline, lastly, Our Final Solution is an ensemble of three models: default nnU-Net, nnU-Net trained with custom TransBTS and our modified TransBTS.} 
\vspace{0.5em}
\begin{tabular}{l|l|l|l|l|l|l}
\Xhline{2\arrayrulewidth}
Method & Dice ET & Dice TC & Dice WT & HD ET & HD TC & HD WT \\
\Xhline{2\arrayrulewidth}
Our TransBTS 500 epochs &  0.78676 & 0.82102 & 0.89721 & 19.826 & 15.1904 & 6.725\\
\hline
Our TransBTS 700 epochs &  0.81912 & 0.82491 & 0.90083 & 15.858 & 16.7709 & 5.8139\\
\hline
nnU-Net default & 0.81536 & \textbf{0.8780} & 0.92505 & 21.288 & 7.76043 & 3.6352\\
\hline
nnU-Net + our TransBTS &  0.84565 & 0.87201 & 0.92394 & 17.364 & 7.78478 & 3.6339\\
\hline
Our TransBTS inside nnU-Net &  0.79818 & 0.86844 & 0.9244 & 24.8750 & \textbf{7.7489} & 3.6186\\
\hline
Our Final Solution & \textbf{0.8496} & 0.86976 & \textbf{0.9256} & \textbf{15.723} & 11.0572 & \textbf{3.3743}\\
\hline
\end{tabular}
\label{comparison_leaderboard}
\end{table} 

\begin{table}[t!]
\centering
\caption{Comparison of TTA techniques on local validation set using our modified TransBTS trained for 500 epochs with self-supervised pretraining. Dice is computed per class, HD corresponds to Hausdorff Distance, W/H/D signifies that three flips were used seperately one dimension flip per TTA component.}
\vspace{0.5em}
\begin{tabular}{l|l|l|l}
\Xhline{2\arrayrulewidth}
TTA Type & Dice NCR & Dice ED & Dice ET \\
\Xhline{2\arrayrulewidth}
w/o TTA &  0.75079 & 0.84068 & 0.82727\\
\hline
All flips \& Rotation & 0.72535 & 0.83284 & 0.81143\\
\hline
W/H/D flips \& Rotation \& Gamma & 0.74752 & \textbf{0.84871} & 0.82896\\
\hline
W/H/D flips & 0.75201 & 0.84297 & 0.82917\\
\hline
W/H/D flips \& Rotation & \textbf{0.75230} & 0.84518 & \textbf{0.83033}\\
\hline
\end{tabular}
\label{tta_experiments}
\end{table}

\begin{table}[t!]
\centering
\caption{Comparison of post processing techniques on challenge validation set. HD stands for Hausdorff Distance metric, while C.C is connected component.}
\vspace{0.5em}
\begin{tabular}{l|l|l|l|l|l|l}
\Xhline{2\arrayrulewidth}
Post Processing Type & Dice ET & Dice TC & Dice WT & HD ET & HD TC & HD WT\\
\Xhline{2\arrayrulewidth}
Original & 0.78676 & 0.82102 & 0.89721 & 19.82 & 15.19 & 6.72\\
\hline
Replacing & 0.81515 & 0.81868 & 0.89403 & 17.74 & 19.84 & 6.95\\
\hline
C.C + Replacing  & 0.81517 & 0.81869 & 0.89406 & 17.75 & 19.84 & 6.96\\
\hline
C.C per class + Replacing & 0.81514 & 0.81868 & 0.89403 & 17.76 & 19.89 & 6.97\\
\hline
\end{tabular}
\label{postproc_experiments}
\end{table}

\par
\begin{figure*}[b!]
\begin{center}
\includegraphics[width=\textwidth]{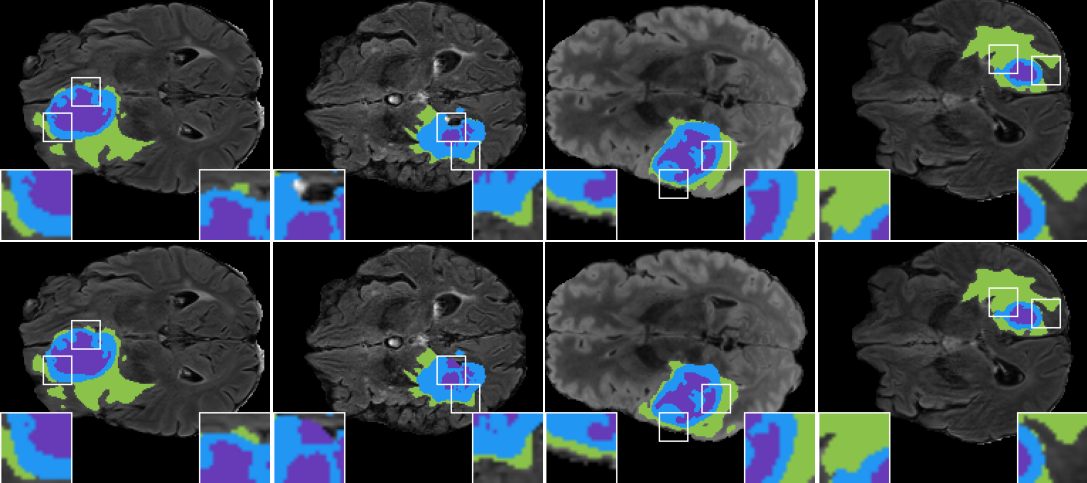}
\end{center}
\vspace{-1.em}
  \caption{Qualitative local validation set results. Upper row represents ground truth masks while lower row contains predictions from our custom TransBTS model trained for 700 epochs.}
\label{fig:qualitative}
\end{figure*}

We show some qualitative results of our segmentation predictions in Figure\ref{fig:qualitative}.

\subsection{GPU Resources}
We implemented our methods in PyTorch\cite{NEURIPS2019_9015} and trained it on a single NVIDIA GeForce RTX 3090 GPU with 24GB graphics RAM size. At inference time we use one NVIDIA GeForce RTX 2080 Ti 11GB GPU. For a model input to fit in these memory constraints we had to decrease the volume size, so we split the whole 3D sample with 240x240x155 dimensionality into 8 smaller overlapping voxels of approximately 128x128x128 pixels and merge the output into one prediction.

%% file: sections/Conclusion.tex
\section{Discussion and Conclusion}
Our proposed solution to BraTS 2021 challenge includes an aggregation of predictions from several models and achieves better performance than any of those methods separately. We selected a CNN with Visual Transformer (based on TransBTS), added customization to its architecture, and incorporated it into the nnU-Net pipeline. This model was used in ensemble with default nnU-Net and our single modified TransBTS. There is still room for improvements in our method and we discuss some ideas below. \par

In our solution, the data augmentations weren't explored in depth. This creates a window of opportunity to improve current results for every model in the proposed ensemble. The easiest way to approach this would be to use augmentations described by winners from last-year challenge \cite{Isensee2021}. \par

% hausdorffERLoss
We also suggest experimenting with computing Hausdorff loss during training and optimizing it alongside Dice and Cross-entropy. This should improve the Hausdorff distance metric and possibly overall performance on dice as well. However, this loss is very time-consuming and is usually implemented on CPU, so we recommended using the version based on applying morphological erosion on the difference between the true and estimated segmentation maps \cite{Karimi2020}, which saves computations.
\par
% didn't use knowledge about classes combination into the whole tumor and tumor core

The knowledge about labels combination into whole tumor and tumor core could be also used during training, perhaps even a separate model trained on most challenging class.

% name of your team? : 'UCU ML lab'
% What is your Team's Synapse ID?     3430761
%  What is the submission ID for validation dataset evaluation?   9715168

%% file: paper.bbl
\begin{thebibliography}{10}
\providecommand{\url}[1]{\texttt{#1}}
\providecommand{\urlprefix}{URL }
\providecommand{\doi}[1]{https://doi.org/#1}

\bibitem{data1}
Baid, U., et~al.: The rsna-asnr-miccai brats 2021 benchmark on brain tumor
  segmentation and radiogenomic classification. arXiv preprint arXiv:2107.02314
   (2021)

\bibitem{data4}
Bakas, S., Akbari, H., Sotiras, A., Bilello, M., Rozycki, M., Kirby, J.,
  et~al.: Segmentation labels for the pre-operative scans of the tcga-gbm
  collection (2017). \doi{10.7937/K9/TCIA.2017.KLXWJJ1Q},
  \url{https://wiki.cancerimagingarchive.net/x/KoZyAQ}

\bibitem{data5}
Bakas, S., Akbari, H., Sotiras, A., Bilello, M., Rozycki, M., Kirby, J.,
  et~al.: Segmentation labels for the pre-operative scans of the tcga-lgg
  collection (2017). \doi{10.7937/K9/TCIA.2017.GJQ7R0EF},
  \url{https://wiki.cancerimagingarchive.net/x/LIZyAQ}

\bibitem{data3}
Bakas, S., Akbari, H., Sotiras, A., Bilello, M., Rozycki, M., Kirby, J.S.,
  et~al.: Advancing the cancer genome atlas glioma {MRI} collections with
  expert segmentation labels and radiomic features. Nature Scientific Data
  \textbf{4}(1) (Sep 2017). \doi{10.1038/sdata.2017.117},
  \url{https://doi.org/10.1038/sdata.2017.117}

\bibitem{Hu2018}
Hu, J., Shen, L., Sun, G.: Squeeze-and-excitation networks. In: 2018
  {IEEE}/{CVF} Conference on Computer Vision and Pattern Recognition. {IEEE}
  (Jun 2018). \doi{10.1109/cvpr.2018.00745},
  \url{https://doi.org/10.1109/cvpr.2018.00745}

\bibitem{Isensee2020}
Isensee, F., Jaeger, P.F., Kohl, S.A.A., Petersen, J., Maier-Hein, K.H.:
  {nnU}-net: a self-configuring method for deep learning-based biomedical image
  segmentation. Nature Methods  \textbf{18}(2),  203--211 (Dec 2020).
  \doi{10.1038/s41592-020-01008-z},
  \url{https://doi.org/10.1038/s41592-020-01008-z}

\bibitem{Isensee2021}
Isensee, F., J\"{a}ger, P.F., Full, P.M., Vollmuth, P., Maier-Hein, K.H.:
  {nnU}-net for brain tumor segmentation pp. 118--132 (2021).
  \doi{10.1007/978-3-030-72087-2\_11}

\bibitem{Isensee2019}
Isensee, F., Petersen, J., Klein, A., Zimmerer, D., Jaeger, P.F., Kohl, S.,
  Wasserthal, J., Koehler, G., Norajitra, T., Wirkert, S., Maier-Hein, K.H.:
  Abstract: {nnU}-net: Self-adapting framework for u-net-based medical image
  segmentation. In: Informatik aktuell, pp. 22--22. Springer Fachmedien
  Wiesbaden (2019). \doi{10.1007/978-3-658-25326-4\_7}

\bibitem{chen2021transunet}
Jieneng, C., Yongyi, L., Qihang, Y., Xiangde, L., Ehsan, A., Yan, W., Le, L.,
  L., Y.A., Yuyin, Z.: Transunet: Transformers make strong encoders for medical
  image segmentation. arXiv preprint arXiv:2102.04306  (2021)

\bibitem{Karimi2020}
Karimi, D., Salcudean, S.E.: Reducing the hausdorff distance in medical image
  segmentation with convolutional neural networks. {IEEE} Transactions on
  Medical Imaging  \textbf{39}(2),  499--513 (Feb 2020).
  \doi{10.1109/tmi.2019.2930068},
  \url{https://doi.org/10.1109/tmi.2019.2930068}

\bibitem{data2}
Menze, B.H., Jakab, A., Bauer, S., Kalpathy-Cramer, J., Farahani, K., Kirby,
  J., et~al.: The multimodal brain tumor image segmentation benchmark
  ({BRATS}). {IEEE} Transactions on Medical Imaging  \textbf{34}(10),
  1993--2024 (Oct 2015). \doi{10.1109/tmi.2014.2377694},
  \url{https://doi.org/10.1109/tmi.2014.2377694}

\bibitem{NEURIPS2019_9015}
Paszke, A., Gross, S., Massa, F., Lerer, A., Bradbury, J., Chanan, G., Killeen,
  T., Lin, Z., Gimelshein, N., Antiga, L., Desmaison, A., Kopf, A., Yang, E.,
  DeVito, Z., Raison, M., Tejani, A., Chilamkurthy, S., Steiner, B., Fang, L.,
  Bai, J., Chintala, S.: Pytorch: An imperative style, high-performance deep
  learning library pp. 8024--8035 (2019),
  \url{http://papers.neurips.cc/paper/9015-pytorch-an-imperative-style-high-performance-deep-learning-library.pdf}

\bibitem{wang2021transbts}
Wang, W., Chen, C., Ding, M., Li, J., Yu, H., Zha, S.: Transbts: Multimodal
  brain tumor segmentation using transformer. In: International Conference on
  Medical Image Computing and Computer Assisted Intervention (MICCAI) (2021)

\bibitem{wang2020modalitypairing}
Wang, Y., Zhang, Y., Hou, F., Liu, Y., Tian, J., Zhong, C., Zhang, Y., He, Z.:
  Modality-pairing learning for brain tumor segmentation  (2020)

\end{thebibliography}
